\def\Gyr{\;{\rm Gyr} }
\def\Myr{\;{\rm Myr} }
\def\yrs{{\rm yrs} }

\def\spose#1{\hbox to 0pt{#1\hss}}
\def\lta{\mathrel{\spose{\lower 3pt\hbox{$\sim$}}
    \raise 2.0pt\hbox{$<$}}}
\def\gta{\mathrel{\spose{\lower 3pt\hbox{$\sim$}}
    \raise 2.0pt\hbox{$>$}}}

\documentstyle[12pt,aasms4,epsf,rotate]{article}

\begin{document}
\title{Possible Long-Lived Asteroid Belts in the Inner Solar System}

\author{N. Wyn Evans and Serge Tabachnik}
\affil{Theoretical Physics, Department of Physics, 1 Keble Rd, Oxford,
OX1 3NP, UK}

\noindent 
{\bf The recent years have witnessed a carnival of discoveries in the
Outer Solar System [1-3].  Here we provide evidence from numerical
simulations of orbital stability to suggest the possible existence of
two long lived belts of primordial planetesimals in the Solar System.
The first is the domain of the Vulcanoids (${\bf \sim 0.09 - 0.21}$
AU) between the Sun and Mercury, where remnant planetesimals may
survive on dynamically stable orbits provided they possess a
characteristic radius greater than ${\bf \sim 0.1}$ km.  The second is
a belt between the Earth and Mars (${\bf \sim 1.08 -1.28}$ AU) on
which an initial population of particles on circular orbits may
survive for the age of the Solar System. A search through the
catalogues of Near-Earth Objects reveals an excess of asteroids with
low eccentricities and inclinations occupying this belt, such as the
recently discovered objects 1996 XB27, 1998 HG49 and 1998 KG3.}
 
Symplectic integrators [4,5] with individual timesteps [6] provide a
fast algorithm that is perfectly suited to long numerical
integrations of low eccentricity orbits in a nearly Keplerian force
field. Individual timesteps are a great boon for our work, as orbital
clocks tick much faster in the Inner Solar System than the Outer. Over
a thousand test particles are distributed on concentric rings with
values of the semimajor axis between $0.1$ AU and $2.2$ AU. Each of
these rings is located in the invariable plane and hosts five test
particles with starting longitudes $n \times 72^\circ$ with
$n=0,\dots,4$. Initially, the inclinations and eccentricities vanish
for the whole sample of test particles. The test particles are
perturbed by the Sun and planets but do not themselves exert any
gravitational forces. The full gravitational effects of all the
planets (except Pluto) are included.  The initial positions and
velocities of the planets, as well as their masses, come from the JPL
Planetary and Lunar Ephemerides DE405.  For all the computations, the
timestep for Mercury is $14.27$ days. The timesteps of the planets are
in the ratio $1:2:2:4:8:8:64:64$ for Mercury moving outward through to
Neptune.  The relative energy error is oscillatory and has a peak
amplitude of $\sim 10^{-6}$ over the 100 million year integration
timespans (c.f. [5,7]). After each timestep, the test particles are
examined. If their orbits have become hyperbolic or have entered the
sphere of influence of any planet [8] or have approached closer than
ten solar radii to the Sun, they are removed from the simulation.
This general procedure is familiar from a number of recent studies on
the stability of test particles in the Solar System [9,10]. For
example, Holman [9] uncovered evidence for a possible belt between
Uranus and Neptune by a similar integration of test particles in the
gravitational field of the Sun and the four giant planets. His
integrations reached the impressive timescale of 4.5 Gyrs -- of the
order of the age of the Solar System. Simulations of the inner Solar
System are much more laborious, as the orbital period of Mercury is
$\sim 88$ days (as compared to $\sim 4332$ days for Jupiter, the giant
planet with the shortest orbital period). This forces us to use a much
smaller timestep and roundoff error becomes a menacing obstacle to
believable results. So, we adopt the strategy of running on a fleet of
nearly twenty personal computers of varying processor speeds, so that
the calculations are performed in long double precision implemented in
hardware. The integration of the orbits of $1050$ test particles for
$100$ Myrs occupied this fleet of computers for over four months.

Fig.~1 shows the results of this calculation. The survival times of
the test particles are plotted against starting semimajor axis. There
are five test particles at each starting position, so the vertical
lines in the figure join five filled circles which mark their ejection
times. The locations of particles that survive for the entire $100$
Myr timespan are marked by diamonds on the upper horizontal
axis. Around each of the terrestrial planets, there is a swathe of
test particles that are ejected rapidly on a precession timescale.
This band is much broader around Mars than the Earth or Venus, perhaps
because of the higher eccentricity of Mars' orbit. There are also
narrow belts of test particles that survive for the full integration.
So, for example, all 50 of the test particles with starting semimajor
axes between $0.1-0.19$ AU are still present at the end of the 100 Myr
integration. The existence of a population of small asteroid-like
bodies -- known as the Vulcanoids -- wandering in intra-Mercurial
orbits has been hypothesised before [11-13].  There are 16 surviving
test particles with starting semimajor axes between $0.6$ and $0.66$
AU, suggesting the possible existence of a narrow belt between Mercury
and Venus.  A somewhat larger third belt of 33 surviving test
particles occupies a belt between Venus and the Earth. Their starting
semimajor axes range from $0.79$ to $0.91$ AU.  Finally, there is a
broad belt between the Earth and Mars from $1.08$ to $1.28$ AU in
which a further 26 test particles survive. The possibility of the
existence of belts between Venus and the Earth and between the Earth
and Mars was raised by Mikkola \& Innanen [10] on the basis of 3 Myr
integrations.

Of course, these results must be treated with considerable reserve, as
100 Myrs is just $\sim 2 \%$ of the age of the Solar System since the
assembly of the terrestrial planets ($\sim 5 \Gyr$).  It is
straightforward to estimate that if this simulation in long double
precision were to be continued till the integration time reaches even
$1 \Gyr$, then it would consume $\sim 3.5$ years of time.
Accordingly, we use the standard, albeit approximate, device of
re-simulating with greater resolution and extrapolating the results.
At semimajor axes separated by $0.002$ AU, five test particles are
again launched on initially circular orbits with starting longitudes
$n \times 72^\circ$ with $n=0,\dots,4$.  The number of test particles
$N(t)$ remaining after time $t$ is monitored for each of the four
belts.  Table 1 gives the results of fitting the data between $1 \Myr$
and $100 \Myr$ to the following logarithmic and power-law decays:
\begin{equation}
N(t) = a + b \log_{10} \Bigl( t [\yrs] \Bigr), 
\qquad\qquad N(t) = {10^c\over \Bigl(t [\yrs]\Bigr)^d}
\end{equation}
In the last two columns, the expected number of test particles
remaining after $1 \Gyr$ and $5 \Gyr$ is computed. The uncertainties
in the fitted parameters suggest that the logarithmic fall-off is a
better -- and more pessimistic -- fit to the asymptotic behaviour than
a power-law (c.f. [9,14]). Our extrapolations suggest that two of the
belts -- those lying between Mercury and Venus, and between Venus and
the Earth -- will become almost entirely depleted after $5
\Gyr$. However, even taking a staunchly pessimistic outlook, it seems 
certain that some of the Vulcanoids and some of the test particles
with starting semimajor axes between $1.08 - 1.28$ AU in the
Earth-Mars belt may survive for the full age of the Solar System.  We
can make a crude estimate of present-day numbers by extrapolation from
the Main Belt asteroids (c.f. [9]). Assuming that the primordial
surface density falls inversely like distance, we find that
the Earth-Mars belt may be occupied by perhaps a thousand or so
remnant objects. Of course, these objects are now outnumbered by the
more recent arrivals, the asteroids ejected via resonances from the
Main Belt, which may number a few thousand in total [15].

A systematic search for Vulcanoids has already been conducted by Leake
et al. [12], who exploited the fact that bodies so close to the Sun
are identifiable from their substantial infrared excess. No candidate
objects were found. However, the survey was limited to a small area of
just 6 square degrees and was estimated to be $\sim 75 \%$ efficient
to detection of bodies brighter than 5th magnitude in the L band. This
result places constraints on the existence of a population of objects
with radii greater than $\sim 50$ km, but minor bodies with the
typical sizes of small asteroids ($\sim 10$ km) will have evaded
detection.  On theoretical grounds, Vulcanoids have been proposed to
resolve apparent contradictions between the geological and the
geophysical evidence on the history of the surface features on Mercury
[11,12].  The robustness of the Vulcanoid orbits partly stems from the
fact that there is only one neighbouring planet and so may be compared
to the stability of the Kuiper-Edgeworth belt. Even after $100 \Myr$, some
$80 \%$ of our Vulcanoid orbits still have eccentricities $e < 0.2$
and inclinations $i < 10^\circ$. It is this evidence, together with
the low rate of attrition of their numbers, that suggests that they
can continue for times of the order the age of the Solar System.  The
outer edge of the Vulcanoid belt is at $\sim 0.21$ AU. Objects beyond
this are dynamically unstable and are excited into Mercury-crossing
orbits on $100 \Myr$ timescales.  The inner edge of the belt is not so
sharply defined. Small objects close to the Sun may be susceptible to
destruction both by Poynting-Robertson drag [16] and by evaporation
[17].  Taking the mean density and Bond Albedo of a typical Vulcanoid
to be the same as that of Mercury, we find that objects with radii
satisfying $ 0.1 \lta z \lta 50$ km can evade both drag and
evaporation in the Vulcanoid belt. This is one of the most dynamically
stable regimes in the entire Solar System. If further searches do not
detect any intra-Mercurial objects, this is a strong indicator that
other processes -- such as planetary migrations -- may have disrupted
the population.

Although there are no known intra-Mercurial bodies, we {\it can} find
candidate objects for the Earth-Mars belt.  Suppose we search an
asteroidal database [18] for objects with inclinations $i <10^\circ$
and eccentricities $e < 0.2$ between the semimajor axes of Earth and
Mars, then we find that there are ten objects.  Of these, seven lie
within our suggested Earth-Mars belt ($1.08 - 1.28$ AU), which is
evidence for an enhancement of nearly circular orbits in this
region. An even more striking test is to search through the objects
between the Earth and Mars for low eccentricity and inclination
asteroids that are not planet-crossing. Then, there are only three
objects (1996 XB27, 1998 HG49 and 1998 KG3) among the entire asteroids
in the database, and all three lie between 1.08 and 1.28 AU.  Most of
the $\sim 50$ asteroids with semimajor axes presently located in our
Earth-Mars belt are moving on orbits with large eccentricities and
inclinations. They are not dynamically stable and will evolve on
timescales of the order of a few Myrs. Most of these objects are
believed to be asteroids ejected from resonance locations in the Main
Belt, although a handful may even be comets whose surfaces have become
denuded of volatiles [19].  However, the seeming enhancement of
circular orbits in this region hints at a primordial population whose
orbits are very mildly eccentric and mildly inclined. Ejection from
the Main Belt will tend to increase the eccentricity of an asteroid
[20]. So, the mildly eccentric objects may well be remnant
planetesimals, the original denizens of the region before it was
colonized by asteroids from the resonance locations in the Main Belt.

\bigskip
\bigskip
\noindent
{\bf Acknowledgments}

\noindent
We thank the Royal Society for the money to purchase dedicated
computers, as well as the Oxford Supercomputing Centre (OSCAR). Above
all, we wish to thank Prasenjit Saha and Scott Tremaine for their
stimulating and insightful suggestions and comments, as well as their
advice on computational matters. Helpful criticism from John Chambers,
Luke Dones and the two referees is also gratefully acknowledged.

\eject

\begin{figure}
\begin{center}
{
              \epsfxsize 0.7\hsize
               \leavevmode\epsffile{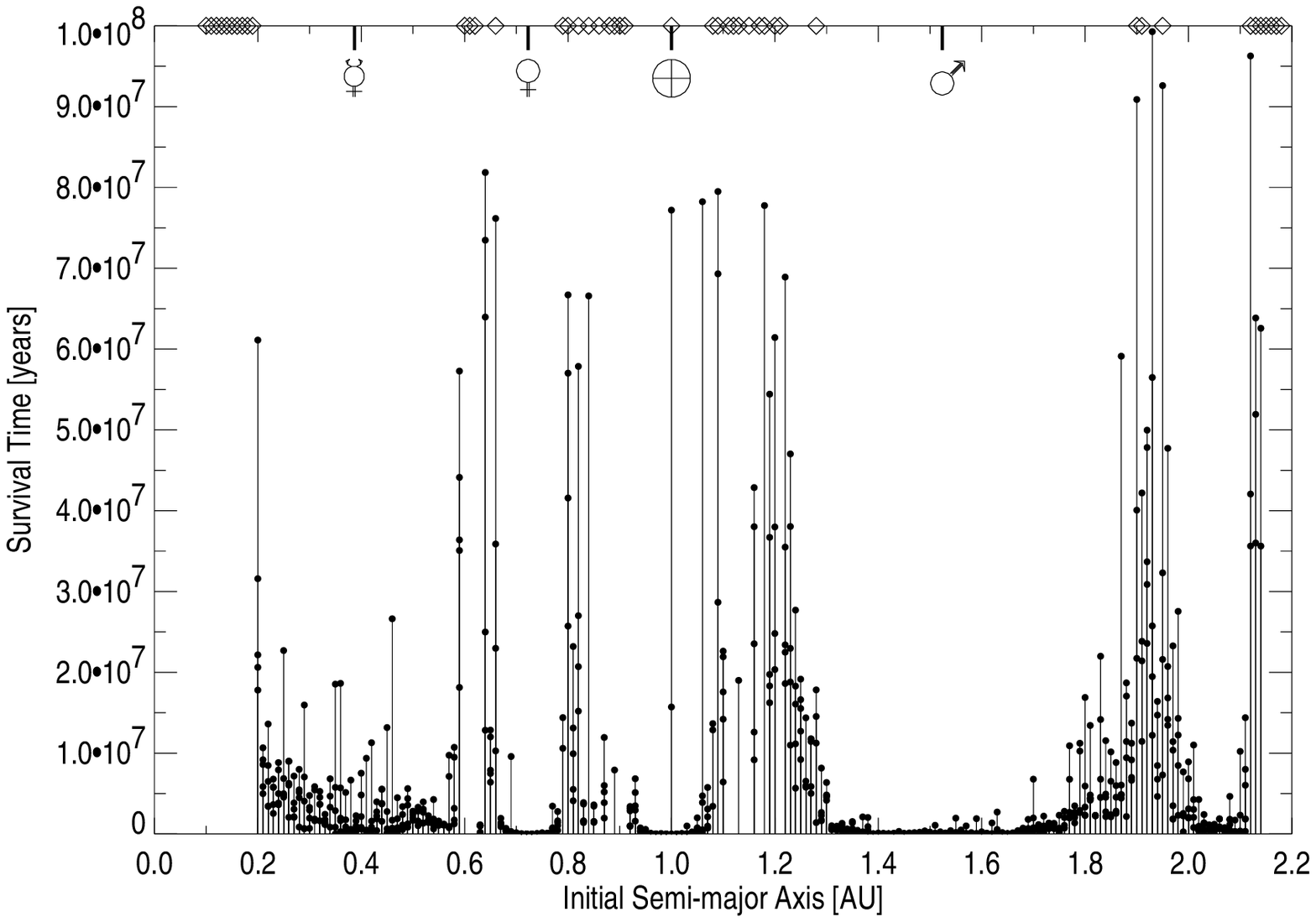}
}
\end{center}
\caption{The survival time (in years) is plotted against starting
semimajor axis (in astronomical units) for test particles in the Inner
Solar System. At each semimajor axis, five test particles are launched
at equally spaced longitudes and their initially circular orbits are
integrated for 100 Myrs. The times of ejection are marked with filled
circles and joined with solid vertical lines. Test particles close to
the terrestrial planets are rapidly removed.  However, between all the
terrestrial planets, there are narrow belts of stable circular orbits
that survive for the full duration of integration of 100 Myrs. The
semimajor axes of the surviving test particles are marked by diamonds
at the top of the figure, together with symbols marking the locations
of the terrestrial planets.  Test particles survive in the following
four regions -- between the Sun and Mercury ($0.1 - 0.19$ AU), between
Mercury and Venus ($0.6 - 0.66$ AU), between Venus and the Earth
($0.79 - 0.91$ AU) and between the Earth and Mars ($1.08 - 1.28$ AU).
The two test particles that remain after $100$ Myrs at an initial
semimajor axis of $1$ AU are actually librating about the Earth's Lagrange
points.  [This calculation has been performed on personal computers
that employ 80 bits internally and which offer the option of
compilation in long double precision with a 64 bit mantissa. At least
64 bits are required to keep Mercury's longitude error below $0.01$
radians over 100 Myr timescales (see [21]).  Standard double precision
offers a mantissa of just 53 bits.]  }
\end{figure}
\eject
\begin{table*}
\begin{center}
\begin{tabular}{rccccc}
Belt & $a$ & $b$ & $N_{\rm 0}$ & $N_{\rm exp}$ (1Gyr) & $N_{\rm exp}$
(5Gyr)\\ \tableline
V & $472.34\pm 5.0$ &$-26.03 \pm 0.68$ & $300$ & $238$ & $220$ \\
M-V & $681.7\pm 3.7$ & $ -69.73 \pm 0.52$ & $250$ & $54$ &$5$ \\
V-E & $1041.1\pm 3.9$  & $-111.51 \pm 0.56$ & $375$ & $38$ & $-$ \\
E-M & $1229.2 \pm 3.2$  & $-118.26 \pm 0.46$ & $500$ &$165$& $82$ \\
\null&\null&\null&\null&\null&\null \\
\end{tabular}
\begin{tabular}{rccccc}
Belt & $c$ & $d$ & $N_{\rm 0}$ & $N_{\rm exp}$ (1Gyr) & $N_{\rm exp}$ 
(5Gyr)\\ \tableline
V & $2.743 \pm 0.008$ & $0.040 \pm 0.001$ & $300$ & $241$ & $226$ \\
M-V & $3.445 \pm 0.017$ & $0.167 \pm 0.002$ & $250$ & $88$ & $67$\\
V-E & $3.725 \pm 0.007$  & $0.189 \pm 0.001$ & $375$ & $106$ & $78$ \\
E-M & $3.531 \pm 0.008$  & $0.133 \pm 0.001$ & $500$ & $215$& $173$ \\
\null&\null&\null&\null&\null&\null \\
\end{tabular}
\end{center}

% Text for table footnotes must follow the tabular environment but must
% be inside the table environment.  Note that it is OK to put \ref's
% in \tablenotetext's.

\tablenum{1}
\caption{The labels V, M-V, V-E and E-M refer to the Vulcanoids 
($0.09-0.21$ AU), the Mercury-Venus belt ($0.58-0.68$ AU), the
Venus-Earth belt ($0.78 -0.93$ AU) and Earth-Mars belt ($1.08 - 1.28$
AU) respectively. Test particles are placed at semimajor axes
separated by $0.002$ AU in each of the belts and the orbits are
re-simulated for 100 million years. The initial number of test
particles in each belt $N_{\rm 0}$ is given.  In each case, the data
$N(t)$ is fitted for $1\Myr < t < 100\Myr$.  The parameters in the
logarithmic and the power-law fits to the number of remaining test
particles $N(t)$ after time $t$ are listed in the upper and lower
tables.  The last two columns of the tables give the extrapolated
number of test particles estimated to remain after 1 Gyr and 5
Gyrs. The estimated uncertainties in the fitted parameters indicate
that the logarithmic law is a better guide for extrapolation -- it is
also more pessimistic than the power-law fit. This suggests that only
the Vulcanoid orbits and the Earth-Mars belt can be accepted as
candidate repositories for long-lived objects. [The simulation of the
Vulcanoid belt has been performed in long double precision on personal
computers, the remaining three belts in double precision on a
supercomputer.]}

\end{table*}


\begin{references}

\reference{}[1] Kowal C.T. in Asteroids (ed Gehrels T.)
p. 436-439 (University of Arizona Press, Tucson, 1979)

\reference{}[2] Jewitt D., Luu J.X. Discovery of the candidate 
Kuiper belt object 1992 QB1, Nature, 362, 730-732 (1993)

\reference{}[3] Williams I.P., O'Cellaigh D.P., Fitzsimmons A., 
Marsden B.G., The slow-moving objects 1993 SB and 1993 SC,
Icarus, 116, 180-185 (1995)

\reference{}[4] Wisdom J., Holman M.J., Symplectic maps for the n-body
problem, Astron. J., 102, 1528-1538 (1991)

\reference{}[5] Saha P., Tremaine S.D., Symplectic integrators for 
solar system dynamics, Astron. J., 104, 1633-1640 (1992)

\reference{}[6] Saha P., Tremaine S.D., Long term planetary
integration with individual time steps, Astron. J., 108, 1962-1969 (1994)

\reference{}[7] Holman M.J., Wisdom J., Dynamical stability in 
the outer solar system and the delivery of short period comets, 
Astron. J., 105, 1987-1999 (1993)

\reference{}[8] Danby J.M.A., Fundamentals of Celestial Mechanics, 
(Willmann-Bell, Richmond, 1988)

\reference{}[9] Holman M.J., A possible long-lived belt of objects
between Uranus and Neptune, Nature, 387, 785-788 (1997)

\reference{}[10] Mikkola S., Innanen K., Solar System Chaos and the
Distribution of Asteroid Orbits, Mon. Not. R. Astron. Soc., 277, 497 (1995)

\reference{}[11] Weidenschilling S.J., Iron/Silicate Fractionation
and the Origin of Mercury, Icarus, 35, 99-111 (1978)

\reference{}[12] Leake M.A., Chapman C.R., Weidenschilling S.J., 
Davis D.R., Greenberg R., The chronology of Mercury's geological and 
geophysical evolution - The Vulcanoid hypothesis, Icarus, 71, 350-375 (1987)

\reference{}[13] Campins H., Davis D.R., Weidenschilling S.J., 
Magee M., in Completing the Inventory of the Solar System,
(eds. Rettig T.W., Hahn J.M.)  p. 85-96, (ASP Conf. Proc.,
Astron. Soc. Pacif., San Francisco, 1996)

\reference{}[14] Dones L., Levison H.F., Duncan M. 1996, in
Completing the Inventory of the Solar System, (eds. Rettig T.W., Hahn
J.M.)  p. 233-244, (ASP Conf. Proc., Astron. Soc. Pacif., San Francisco,
1996)

\reference{}[15] Weissman P.R, A'Hearn M.F., McFadden L.A.,
Rickman H., in Asteroids II, (eds. Binzel R.P., Gehrels T., 
Matthews M.S.) p. 880-920, (University of Arizona Press, Tucson,
1989)

\reference{}[16] Robertson H.P., Dynamical effects of radiation in
the solar system, Mon. Not. R. Astron. Soc., 97, 423-438 (1937) 

\reference{}[17] Pettit E., in Planets and Satellites, (eds. Kuiper
G.P., Middlehurst B.) p. 400-427, (University of Chicago Press,
Chicago, 1961)

\reference{}[18] Bowell E.G., The Asteroid Orbital Element
Database, at ftp://ftp.lowell.edu/pub/elgb/astorb.html

\reference{}[19] McFadden L.A., Tholen D.J., Veeder G.J., in 
Asteroids II, (eds. Binzel R.P., Gehrels T., Matthews M.S.)
p. 442-467, (University of Arizona Press, Tucson, 1989)

\reference{}[20] Gladman B. et al. Dynamical Lifetimes of
Objects Injected into Asteroid Belt Resonances, Science, 277,
197-201 (1997)

\reference{}[21] Saha P., Stadel J., Tremaine S.D., A
Parallel Integration Method for Solar System Dynamics, Astron. J., 
114, 409-414 (1997)

\end{references}
\end{document}